\documentclass[twocolumn]{aastex631}
\usepackage{comment}
\usepackage{natbib} 
\usepackage{amsmath}
\usepackage[flushleft]{threeparttable}

\newcommand{\degree}{$^{\circ}$}

\graphicspath{{./}{figures/}}

\defcitealias{Butterfield2024a}{FIREPLACE I}
\defcitealias{Butterfield2024b}{FIREPLACE II}
\defcitealias{Pare2024}{FIREPLACE III}

\begin{document}

\title[PAH III (Galactic Plane)]{\uppercase{Analyzing PAHs as a Tracer of Anomalous Microwave Emission Near the Galactic Plane Using the COSMOGLOBE DIRBE Reduction}}

\correspondingauthor{Dylan M. Par\'e}
\email{dylanpare@gmail.com}

\author[0000-0002-5811-0136]{Dylan M. Par\'e}
\affiliation{Joint ALMA Observatory, Alonso de Cordova 3107, Vitacura, Casilla 19001, Santiago de Chile, Chile}
\affiliation{National Radio Astronomy Observatory, 520 Edgemont Road, Charlottesville, VA 22903, USA}

\author[0000-0003-0016-0533]{David T. Chuss}
\affiliation{Department of Physics, Villanova University, 800 E. Lancaster Ave., Villanova, PA 19085, USA}

\author[0000-0003-3922-1487]{Danielle Sponseller}
\affiliation{Department of Space, Earth and Environment, Chalmers University of Technology, Gothenburg, Sweden}

\author[0000-0001-7449-4638]{Brandon Hensley}
\affiliation{Jet Propulsion Laboratory, California Institute of Technology, 4800 Oak Grove Drive, Pasadena, CA 91109, USA}

\author[0000-0001-9835-2351]{Alan Kogut}
\affiliation{Code 665, Goddard Space Flight Center, Greenbelt, MD 20771, USA}

\begin{abstract}

The physical mechanism producing Anomalous Microwave Emission (AME) has been an unresolved puzzle for close to 30 years.  One candidate mechanism is rotational emission from polycyclic aromatic hydrocarbons (PAHs) which can have the necessary electric dipole moment and size distribution to account for the AME in representative interstellar environments. However, previous investigations have found that AME is better correlated with the far-infrared dust emission rather than the PAH emission. In this work we analyze the correlations between the AME and the PAH and far-infrared dust emission using the 3.3 $\mu$m PAH emission feature as observed by band 3 of the Diffuse Infrared Background Experiment (DIRBE). This analysis builds on previous work conducted in individual molecular clouds and extends it into fainter, more diffuse structures. In addition, we utilize the COSMOGLOBE DIRBE reduction for this work, building on previous studies that used the original DIRBE data set. We find that the AME is better correlated with far-infrared dust emission ($\rho\sim$0.9) than the PAH emission ($\rho\sim$0.7) in the central $|b|\leq$ 10\degree\ region of the sky. This could indicate either that non-PAH dust grains or an alternative physical emission mechanism is primarily responsible for the AME in the Galactic Plane, or that the excitation conditions for mid-infrared emission and for AME from PAHs differ substantially.
 
\end{abstract}


\section{INTRODUCTION} \label{sec:intro}
Anomalous Microwave Emission (AME) has been a known phenomenon for almost 30 years \citep{Kogut1996,Leitch1997}. Despite many investigations into the nature of AME, it remains unclear what emission mechanism is primarily responsible for generating it \citep[see][for a review]{Dickinson2018}. 

One possibility for how AME is generated is rotational emission from polycyclic aromatic hydrocarbons (PAHs). Small, rapidly spinning dust grains with a nonzero electric dipole moment generate electric dipole radiation \citep[the ``spinning dust'' model, e.g.][]{Draine1998,Hoang2011}. PAHs are a species of grain known to exist throughout the interstellar medium (ISM) that have both the necessary dipole moment from carbon substitutions and the small grain sizes needed to account for the observed AME \citep{Draine1998}.This combination of PAH properties makes them a compelling candidate emission mechanism for AME. We note, however, that any small grain with a dipole moment would be capable of producing the AME \citep{Hoang2016a, Hoang2016b, Hensley2017}.

There have been multiple efforts to study spatial correlations between PAHs and AME \citep[e.g.,][]{Ysard2010,Tibbs2011,Planck-collaboration2014b,Bell2019}. Though the understanding of this correlation is mixed, there are indications that AME is better correlated with the dust emission than with the PAH emission. \citet{Hensley2016}, for example, inspected PAH emission throughout the sky as probed by the Wide-field Infrared Survey Explorer (WISE). They used the W3 band centered on a wavelength of 12~$\mu$m to conduct a correlation analysis between the ratio of the W3 emission and FIR dust radiance ($f_{PAH}$) and the AME. They found that the AME is better correlated with the far-infrared thermal dust emission than the WISE data, indicating that thermal dust emission is a better tracer of AME than PAHs. More recently, \citet{Chuss2022} found similar results in $\lambda$-Orionis using DIRBE band 3 to study the PAH 3.3 $\mu$m emission feature. Furthermore, \citet{Sponseller2025} found that far-infrared dust emission is a better tracer of AME than the PAH 3.3 $\mu$m emission feature derived from the Diffuse Infrared Background Experiment (DIRBE) observations in the majority of the 98 prominent AME sources they studied. These recent studies corroborate the results of \citet{Hensley2016}, finding that thermal emission from large dust grains tends to be a better tracer of AME than PAHs. It should be noted, however, that \citet{Sponseller2025} did find 17 regions where PAH emission is a better tracer of AME than the dust emission, indicating that the picture may be more complicated.

The lower correlation between the PAH emission and AME could indicate a few possibilities. Spinning nano-silicate grains that are not PAHs could be the primary generators of AME \citep{Hoang2016a,Hensley2017}. PAHs could be the AME carriers, but since the PAH emission physics and AME have different dependencies on local interstellar conditions the observed correlation between them could be eroded \citep{Hensley2022,Ysard2022}. Alternatively, AME may not be generated by spinning dust emission at all, but could instead originate from thermal vibrational emission \citep{Jones:2009, Nashimoto:2020}.

In this work we build on previous correlation efforts, such as those done by \citet{Sponseller2025}, by studying the AME correlations with both the PAH emission and far-infrared dust emission over the entire Galactic Plane. To do so, we utilize the COSMOGLOBE reduction of the DIRBE data products that allow for better constraints of foreground contaminants \citep[like Zodiacal and thermal dust emission,][]{Watts2023,San2024}. We use the DIRBE data to create maps of the 3.3 $\mu$m PAH emission feature throughout the sky, which traces the smallest PAH grains -- those that are most likely to be producing AME \citep{Draine:2001}. We then correlate the derived PAH emission with the AME and thermal dust emission, quantitatively assessing these correlations over the entire Galactic Plane for the first time.

In Section~\ref{sec:data} we discuss the data sets used and in Section~\ref{sec:meth} we describe the procedure used to generate the maps of the PAH 3.3~$\mu$m feature from the COSMOGLOBE DIRBE data. We present the AME correlations in Section~\ref{sec:res}, and discuss the correlations in Section~\ref{sec:disc}. We present our conclusions in Section~\ref{sec:conc}. 

\section{DATA} \label{sec:data}
\begin{figure*}
    \centering
    \includegraphics[width=1.0\textwidth]{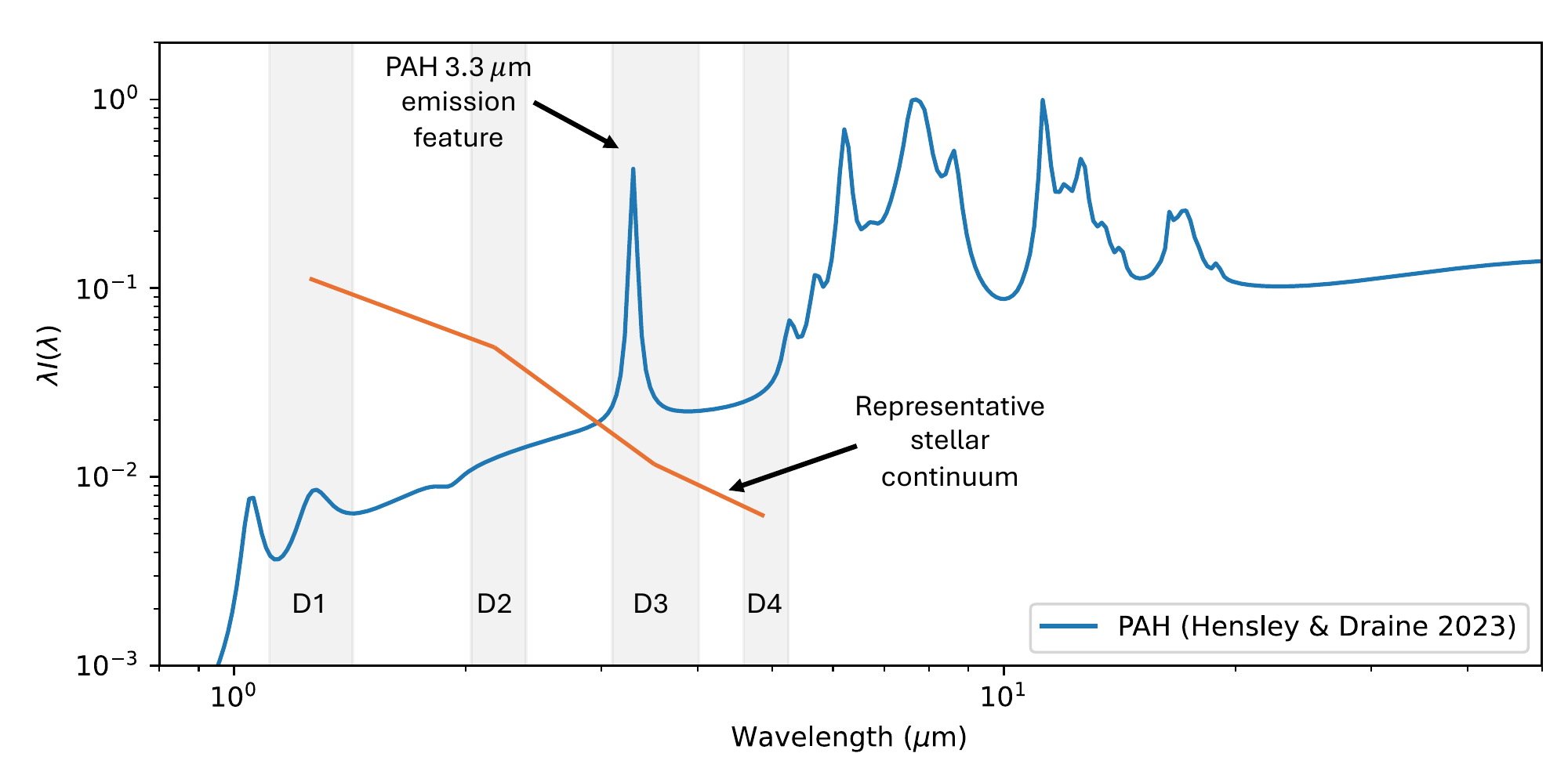}
    \caption{The PAH spectrum as modeled in \citet{Hensley2023} is shown in blue. Grey rectangles indicate the wavelength extents of DIRBE bands 1 -- 4 as marked. The PAH 3.3 $\mu$m emission feature due to the C-H stretching mode in band 3 is also marked. We note that bands 1 and 2 are dominated by the stellar continuum, which we indicate with the orange line.}
    \label{fig:pah_spec}
\end{figure*}
\begin{figure*}
    \centering
    \includegraphics[width=0.49\textwidth]{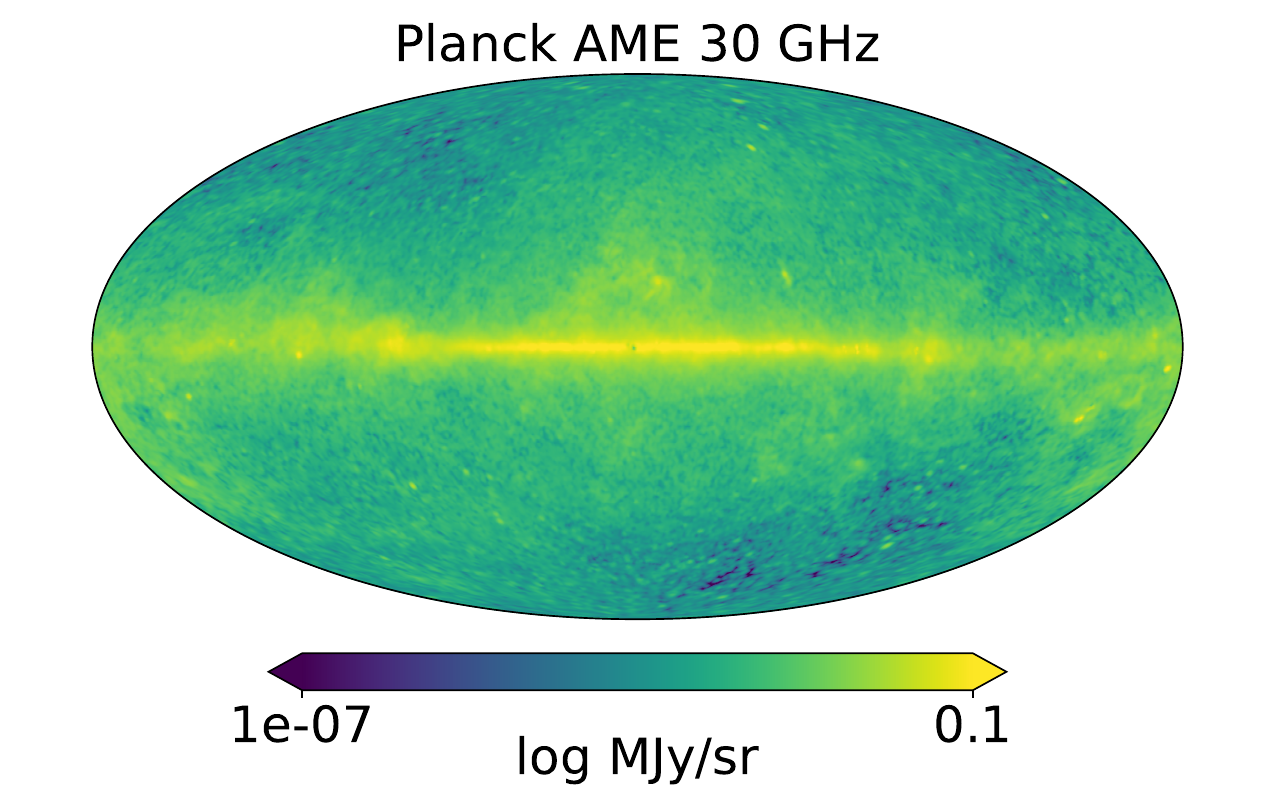}
    \includegraphics[width=0.49\textwidth]{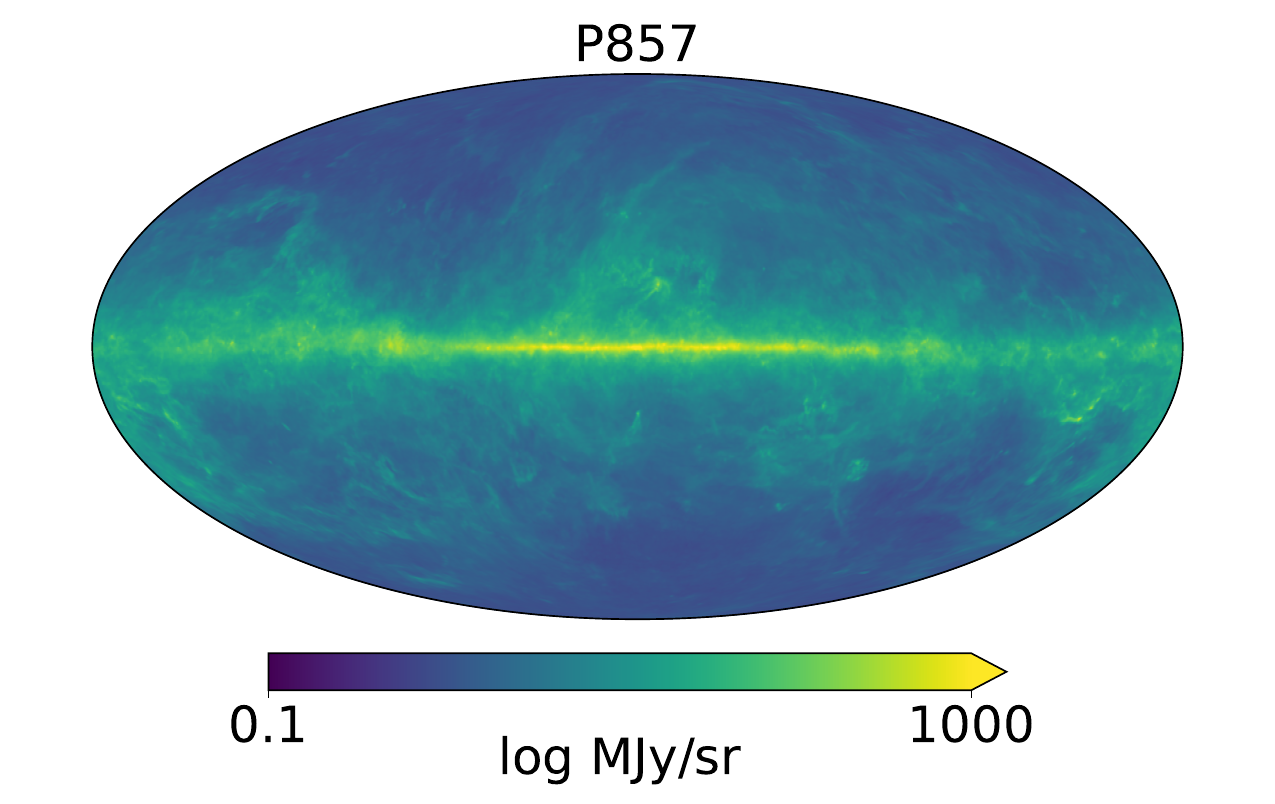}\\
    \includegraphics[width=0.49\textwidth]{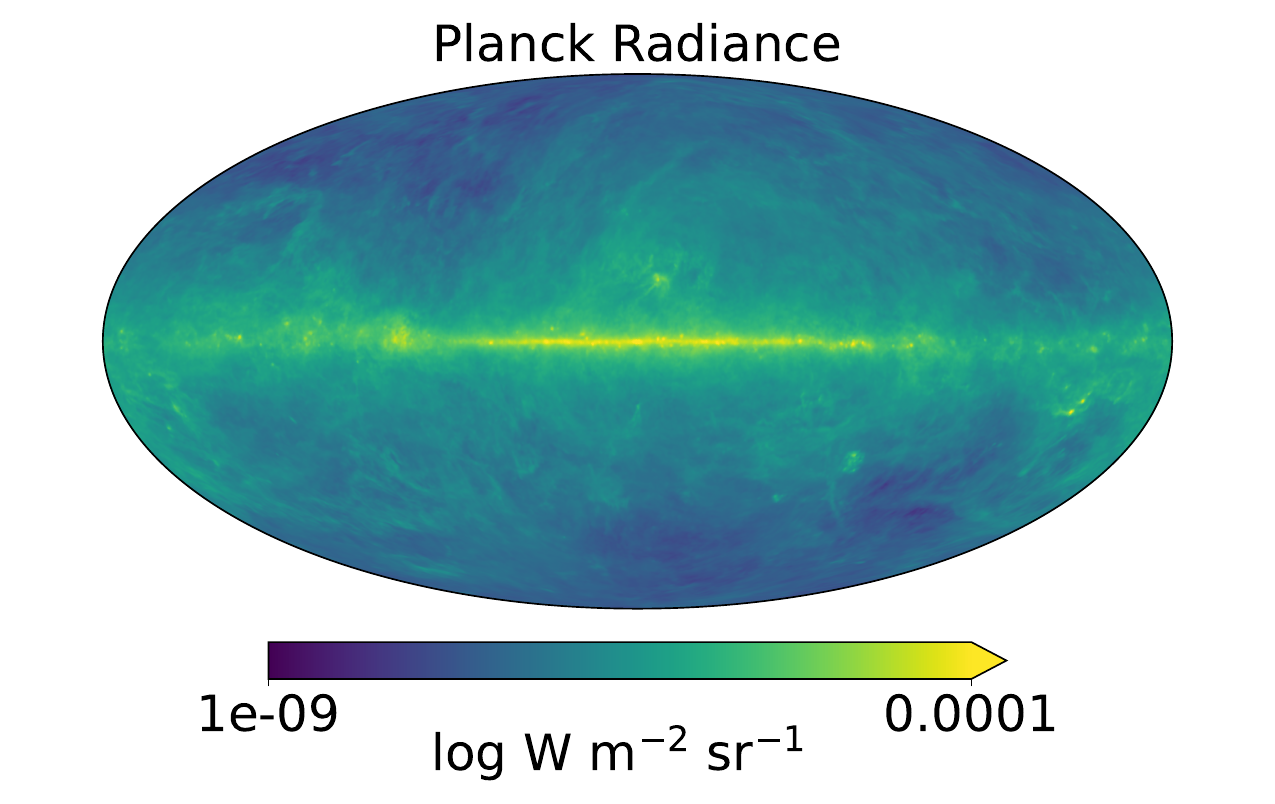}
    \caption{Upper left: The 2-component Commander AME model evaluated at 30 GHz. Upper right: Planck 857 GHz dust map. Bottom: The Planck radiance map with an $N_{\rm side}$ of 512. All maps are smoothed to a 0.5\degree\ resolution and are shown in a logarithmic color scale.}
    \label{fig:AME-dust-R}
\end{figure*}
\begin{figure*}
    \centering
    \includegraphics[width=0.95\textwidth]{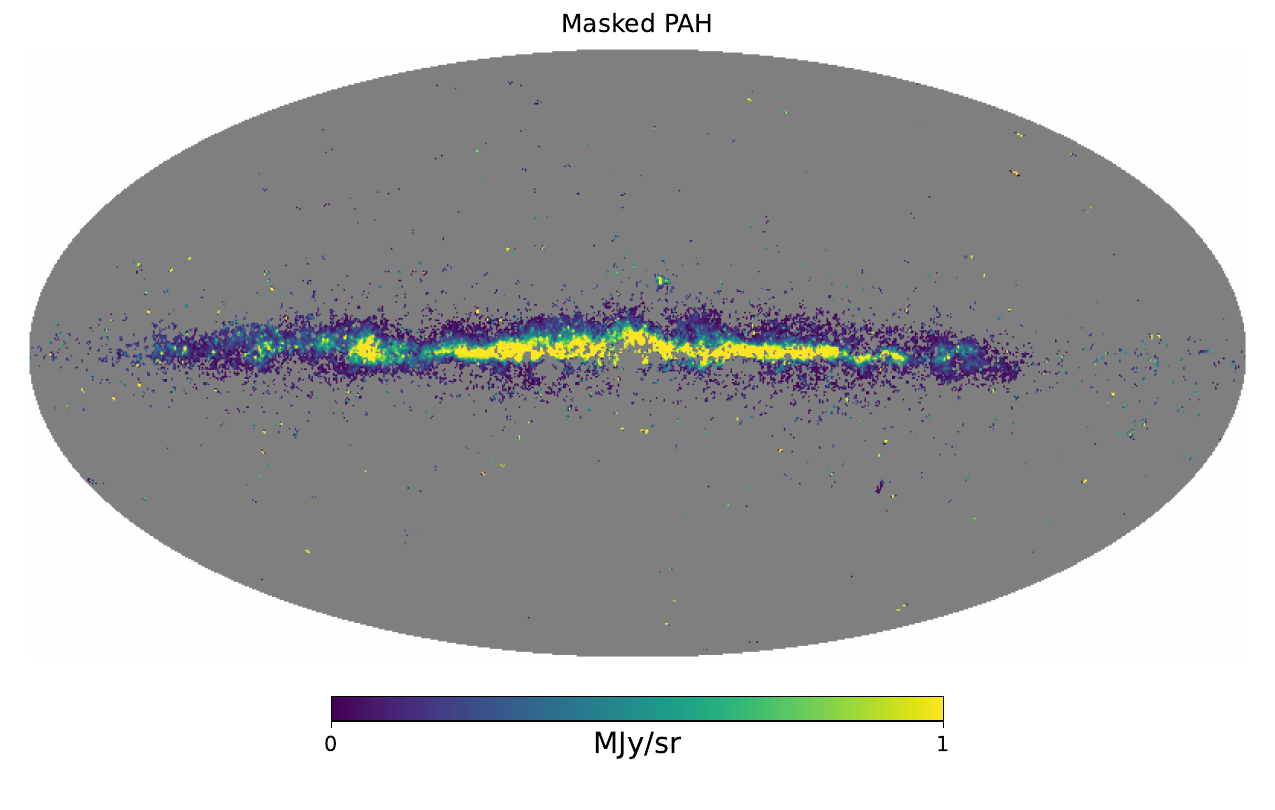}
    \includegraphics[width=0.95\textwidth]{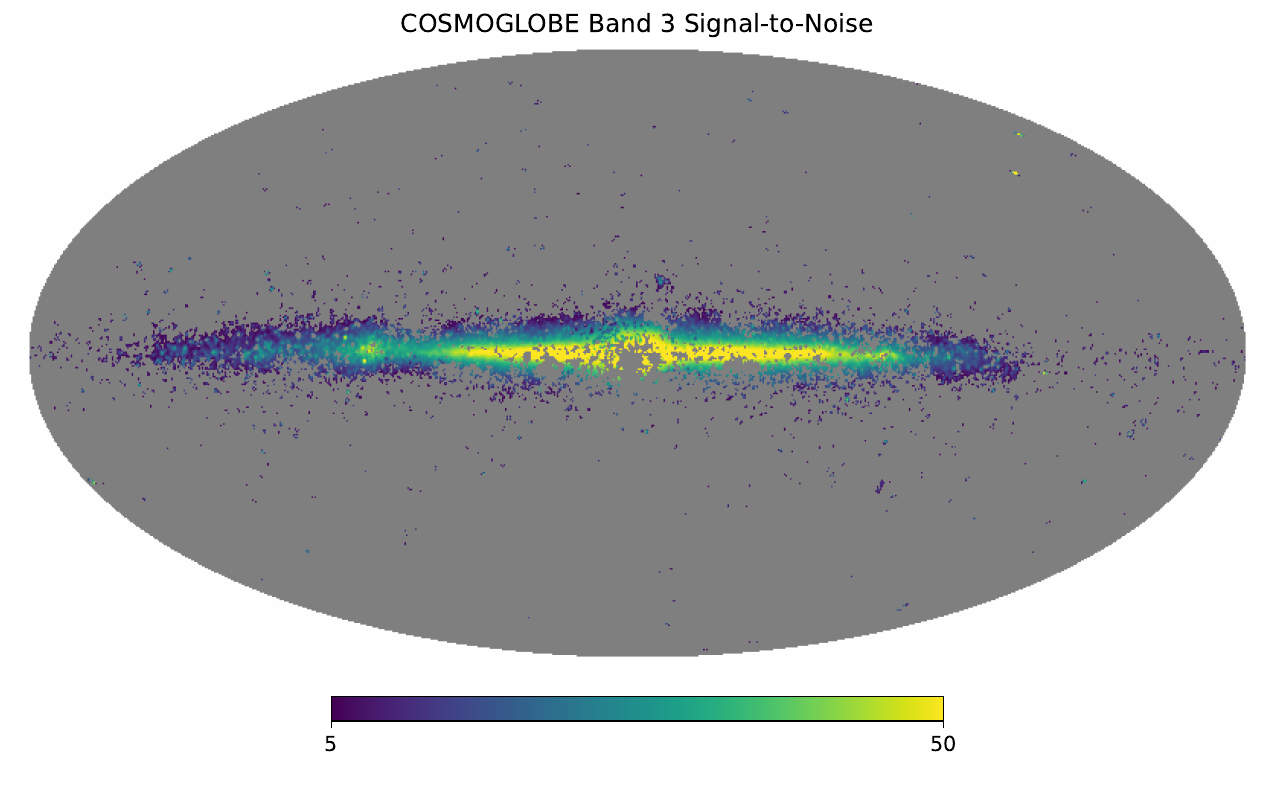}
    \caption{Top: 3.3 $\mu$m PAH emission feature derived from the COSMOGLOBE DIRBE data products and smoothed to a 0.5\degree\ resolution. The map is shown in a linear scale with bright stellar sources and PAH emission coinciding with DIRBE signal-to-noise $<$ 5 masked out, as indicated by the gray points in the image. Bottom: the signal-to-noise ratio of COSMOGLOBE DIRBE band 3 used to mask the PAH map, smoothed to a 0.5\degree\ resolution and shown in a logarithmic scale.  The mask applied to the PAH emission has also been applied to the displayed signal-to-noise map.}
    \label{fig:PAH}
\end{figure*}
To derive a map of the PAH emission, we utilize the COSMOGLOBE reduction of the Cosmic Background Explorer (COBE) satellite DIRBE full-sky maps covering bands 1 -- 4 (at central wavelengths of 1.25, 2.5, 3.5, and 4.9 $\mu$m), which are publicly available\footnote{https://www.cosmoglobe.uio.no/products/cosmoglobe-dr2.html} \citep{Watts2023}. The third DIRBE band (with a central wavelength of 3.5 $\mu$m) contains a prominent PAH emission feature at 3.3 $\mu$m, which occurs as the result of C-H stretching modes. An example PAH spectrum is shown in Figure~\ref{fig:pah_spec} \citep{Hensley2023}, with the frequency widths of the first four DIRBE bands also indicated and the 3.3~$\mu$m PAH emission feature labeled. We also show the stellar continuum emission over the first four DIRBE bands. The 3.3~$\mu$m emission feature is generated by the smallest PAH grains \citep{Draine:2001} and it is fully contained within the DIRBE band 3 wavelength coverage as shown in Figure~\ref{fig:pah_spec}.

We note that the DIRBE observations are composed of 10 bands with wavelengths ranging from 1.25 to 240 $\mu$m; however, we do not include bands 5 -- 10 in this analysis since the emission from these bands is dominated by the thermal continuum emission from larger grains. 

The PAH emission derived from the COSMOGLOBE data set is then compared to the AME and dust emission for the correlation. For the AME we use the spectral template of the Planck observations \citep{Planck-Collaboration2016a}, which we evaluate at 30~GHz. The resulting AME map is smoothed to a resolution of 0.5\degree\ and is shown in the upper-left panel of Figure \ref{fig:AME-dust-R}. For the thermal dust emission we use the Planck Public Release 2 (PR2) 857~GHz map, which we show in the upper-right panel of Figure \ref{fig:AME-dust-R}.

We also compare the PAH emission and AME to the Planck radiance map, which we obtain from the Planck PR2 foreground maps available on IPAC\footnote{https://irsa.ipac.caltech.edu/data/Planck/release\_2/all-sky-maps/foregrounds.html} \citep{Planck2016_fore}. This radiance map is shown in the bottom panel of Figure~\ref{fig:AME-dust-R}.

\section{METHODS} \label{sec:meth}
\subsection{Creating the Milky Way PAH Map}
In order to conduct a meaningful correlation analysis it is necessary to mask out significant emission contaminants in the DIRBE data. Since the COSMOGLOBE collaboration has already performed an extensive analysis (and correction) of the Zodiacal emission for the four DIRBE bands \citep{San2024}, we can perform a spectral fit of the starlight and PAH emission feature allowing us to employ a modified version of the method used by \citet{Sponseller2025} to isolate the PAH emission feature for the subsequent correlation analysis. 

Bright stellar sources are masked out of the COSMOGLOBE DIRBE observations by first calculating the background level in band 1 using an annular background region of radius 0.75 -- 1.00 degree around each pixel. Note that the DIRBE beam size is 0.75\degree, meaning that the lower bound of the annulus is larger than the DIRBE beam size.  All pixels with intensity $\geq$0.5 MJy sr$^{-1}$ above this local background level are then masked. This results in $\ll$1\% of the sky being masked.

With the bright stellar sources so masked, we then perform a spectral fit of the residual starlight and PAH emission using COSMOGLOBE DIRBE bands 1 -- 4. Amplitudes for these components are derived using the linear least-squared fitting (LLS) technique as described below.

The starlight component is modeled using the Faint Source Model (FSM), which is a statistical model that accounts for unresolved starlight and is designed for analysis of DIRBE observations \citep{Arendt1998}. The FSM accounts for the distribution of stars in the Galaxy, the amount of reddening experienced by the stars as a result of dust, and the composition of the stellar population. This model provides a predicted spectrum resulting from the starlight for each line-of-sight. A representative spectrum derived from this model for our observations is shown as the orange line in Figure~\ref{fig:pah_spec}.

To relate the FSM model to the COSMOGLOBE DIRBE observations we use the nearest pixel to the HEALPix\footnote{https://sourceforge.net/projects/healpix/} representation of the COSMOGLOBE DIRBE data set \citep{Gorski2005}. For this work, we chose a number of sides $N_{side} = 512$, corresponding to a pixel size of $\sim$0.1\degree.

To isolate the PAH emission feature, we use the LLS method to fit the spectral basis functions of the PAH and FSM emission components for each HEALPix pixel in the COSMOGLOBE DIRBE data set. We use $X_1(\nu)$ and $X_2(\nu)$ to represent the basis functions for the PAH and FSM emission, respectively. We then construct a model, $Y(\nu)$, that is the linear combination of these basis functions such that for each map pixel
\begin{equation}
    Y(\nu_i) = a_1\times{}X_1(\nu_i) + a_2\times{}X_2(\nu_i), \label{eq:Y}
\end{equation}
where $\nu_i$ is one of the four DIRBE band frequencies used in this work and $a_1$ and $a_2$ are the amplitudes for the PAH and FSM emission components at that frequency which we are solving for using this model. A model solution $Y$ is derived for each pixel of the COSMOGLOBE DIRBE data set. This is the same LLS procedure employed in \citet{Sponseller2025}, except that we do not include a basis function for the Zodiacal emission since the COSMOGLOBE data have already been corrected for this component \citep{San2024}.

To solve the model, $Y$, for the parameters, $a_1, a_2$, we first define a goodness-of-fit metric $X^2$:
\begin{equation}
    X^2 = \sum_{i=1}^4\left[\frac{y_i - Y(\nu_i)}{\sigma_i}\right]^2 \label{eq:x2}
\end{equation}
where $y_i$ is the observed intensity at frequency $\nu_i$ for DIRBE bands 1 -- 4 and $\sigma_i$ is the measurement noise at that band. We then find $a_1,a_2$ such that Equation~\ref{eq:x2} is minimized. The derived PAH emission is composed of both positive and negative values, where the negative values originate from low signal-to-noise regions of the DIRBE map. We implement a signal-to-noise cut of 5 using the COSMOGLOBE band 3 data. The resulting PAH 3.3 $\mu$m map obtained by performing the LLS minimization and masking bright stellar sources and implementing the signal-to-noise cut, is shown in the top panel of Figure \ref{fig:PAH}. We also display the signal-to-noise map of COSMOGLOBE band 3 that remains after implementing the masking in the bottom panel of Figure \ref{fig:PAH}. 

We test whether we are able to recover more of the higher latitude PAH emission by smoothing the PAH and COSMOGLOBE maps to a $\sim$1\degree\ resolution (equivalent to $N_{side} = 64$) after masking the bright stellar sources but before applying the signal-to-noise cut. We find that this lower resolution does not provide any additional insight to the higher latitude regime off the Galactic plane ($\ll$1\% increase in usable sky fraction). We therefore analyze the 0.5\degree\ resolution map presented in Figure \ref{fig:PAH} for all subsequent analysis.

To verify the robustness of the 2-component model fit used to derive the PAH emission presented in Figure \ref{fig:PAH} we display the coefficient for the FSM component of the model fit (corresponding to $a_2$ in Equation \ref{eq:Y}), which we display in Figure \ref{fig:FSM} in Appendix \ref{sec:appendix}. We also show the covariance between $a_1$ and $a_2$ components in Figure \ref{fig:covar} in Appendix \ref{sec:appendix}. The distribution is close to 0 throughout the sky, indicating that there is no significant relationship between the PAH and FSM model components.

We also verified the COSMOGLOBE correction for the Zodiacal emission by performing a 3-term fit to the COSMOGLOBE observations using the same methodology as \citet{Sponseller2025}, where the additional third term accounts for the Zodiacal emission. The Zodiacal emission for each pixel was computed from the Interplanetary Dust (IPD) model presented in \citet{Kelsall1998}, where for each skycube pixel the average of all weekly IPD model predictions was used. This skycube map was then converted to a HEALPix map using the nearest skycube pixel.  We found that the incorporation of this Zodiacal term did not significantly change the resulting PAH map, and so we use the PAH map generated from the 2-component model fit to produce the correlation results presented below.

\section{RESULTS} \label{sec:res}
\subsection{PAH-AME Galactic Plane Correlations}
\begin{figure*}
    \centering
    \includegraphics[width=1.0\textwidth]{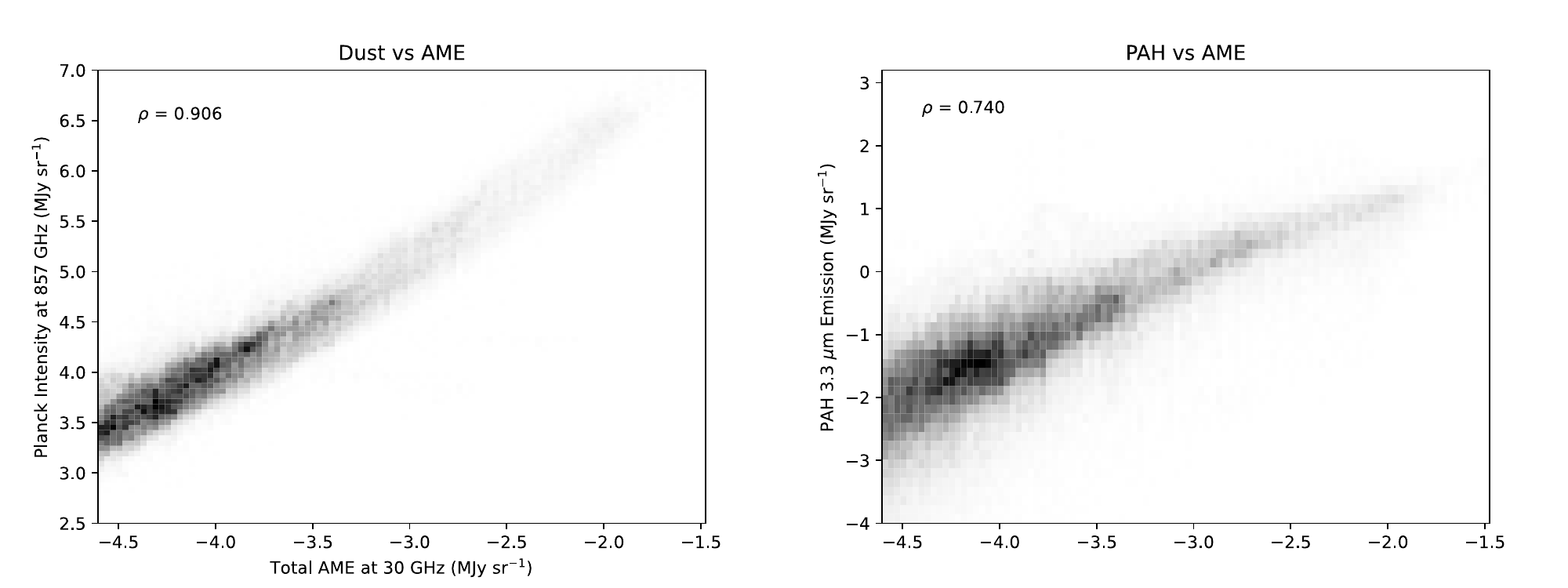}
    \caption{Galactic Plane correlations between the dust, PAH, and AME. Left: correlation between the AME and the dust emission in log-log space. Right: correlation between the AME and the PAH 3.3 $\mu$m emission feature in log-log space. The Spearman-r rank correlation coefficient is shown in the upper left of both panels.}
    \label{fig:corr_full}
\end{figure*}
Galactic Plane correlations of the PAH emission and the AME is shown in the right-hand panel of Figure \ref{fig:corr_full}. The correlation is presented as a 2D-histogram in log-log space. The left-hand panel of Figure \ref{fig:corr_full} presents the 2D correlation of the dust emission and AME in the same format as the right-hand panel. The AME is more correlated with the dust emission $\rho=0.906$) than the PAH emission ($\rho=0.740$), as can be seen from the Spearman-r rank correlation coefficients for the correlations which are shown in the panels. We note that the PAH map is strongly correlated with both the dust emission and the AME map, an indication that the modeling of the PAH coefficient was accurate. To rule out possible SNR effects we study how the Spearman-r correlation coefficient varies as a function of SNR. We find that the coefficient does not strongly depend on the SNR threshold, indicating that the difference in correlation is likely a physical difference rather than an error in the modeling.

\begin{figure*}
    \centering
    \includegraphics[width=1.0\textwidth]{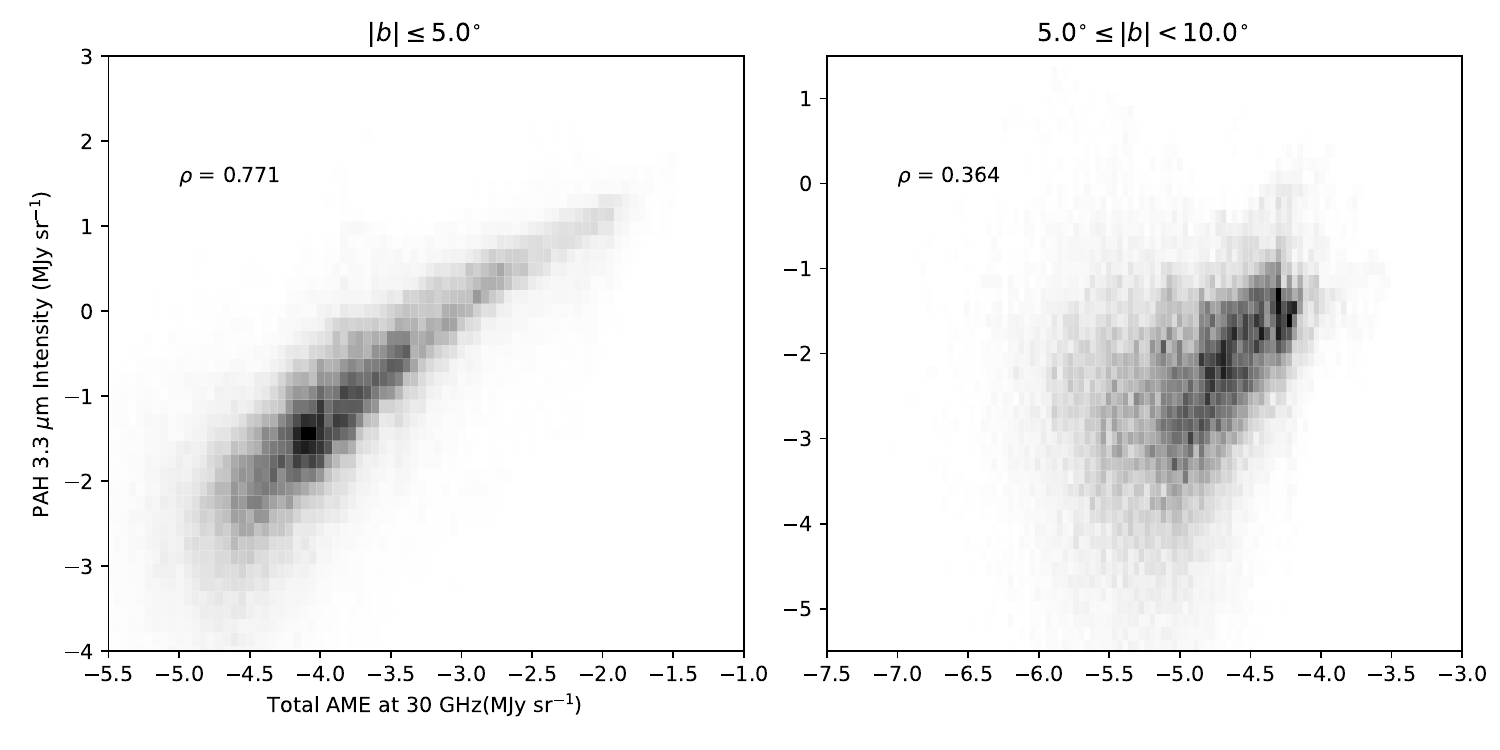}
    \caption{Correlations of the PAH emission and the AME in 5\degree\ wedges of the sky as described in the text. Each histogram is presented in the same way as the right-hand panel of Figure \ref{fig:corr_full}. The Spearman-r rank correlation coefficient is also shown in each panel.}
    \label{fig:PAH_corr}
\end{figure*}
\begin{figure*}
    \centering
    \includegraphics[width=1.0\textwidth]{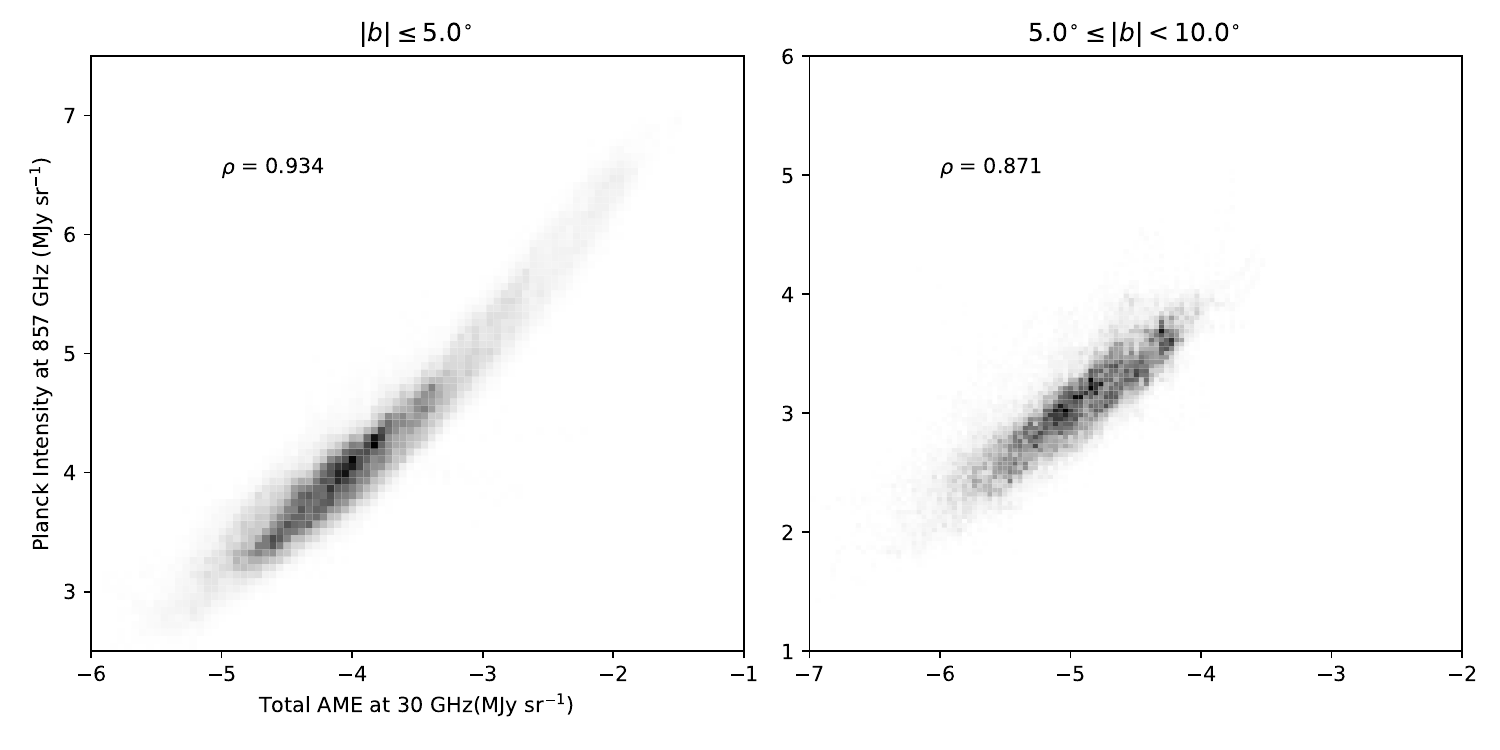}
    \caption{Correlations of the dust emission and the AME in 5\degree\ wedges of the sky as described in the text. Each histogram is presented in the same way as the left-hand panel of Figure \ref{fig:corr_full}. The Spearman-r rank correlation coefficient is also shown in each panel.}
    \label{fig:dust_corr}
\end{figure*}
\begin{deluxetable}{lcc}
\tablecaption{Preferential Alignment with Latitude
\label{tab:alignment}}
\tablewidth{0pt}
\tablehead{
\colhead{Aligment Parameter} & \colhead{$|b| <$ 5\degree} & \colhead{5\degree\ $\leq |b| \leq$ 10\degree}}
\startdata
$\rm\rho_{dust}$   & 0.93 & 0.87 \\
$\rm\rho_{PAH}$   & 0.77 &  0.36 \\
$\rm\eta_{pref}$  & -111 & -128    \\
\hline
\enddata
\tablecomments{For each row the first column indicates the name of the alignment parameter (Spearman-r rank correlation and preferential orientation as defined in Equation \ref{eq:pref}), the second and third columns indicate the value of each parameter for the latitude wedges from 0\degree\ $\leq |b| \leq$ 5\degree\ and 5\degree\ $\leq |b| \leq$ 10\degree, respectively.}
\end{deluxetable}
We also test the dependence of the correlations on Galactic latitude.  To do so, we analyze 5\degree\ wedges of the sky from 0\degree\ $\leq |b| \leq$ 5\degree\ and from 5\degree $\leq |b| \leq$ 10\degree. For higher latitudes ($|b| \geq 10.0$) we do not obtain many PAH detections from the signal-to-noise cut (as can be seen in Figure \ref{fig:PAH}), and so we do not evaluate correlations for these higher latitude regimes.

The correlations between the PAH and dust emission with the AME are displayed in Figures \ref{fig:PAH_corr} and \ref{fig:dust_corr}, respectively. Both panels in these figures also display the the Spearman-r rank correlation coefficient for the correlation.

The AME is more correlated with the dust emission than the PAH emission in the latitude wedges, as indicated by the coefficient values shown in Figures \ref{fig:PAH_corr} and \ref{fig:dust_corr}. To further visualize this trend we present the Spearman-r correlation coefficients obtained from the latitude wedges in the top two rows of Table \ref{tab:alignment}. It can be seen from this table that the Spearman-r coefficient is larger for the AME and dust correlation than it is for the AME and PAH correlation for both latitude wedges.

\subsection{Comparison to Dust Radiance}
\begin{figure*}
    \centering
    \includegraphics[width=1.0\textwidth]{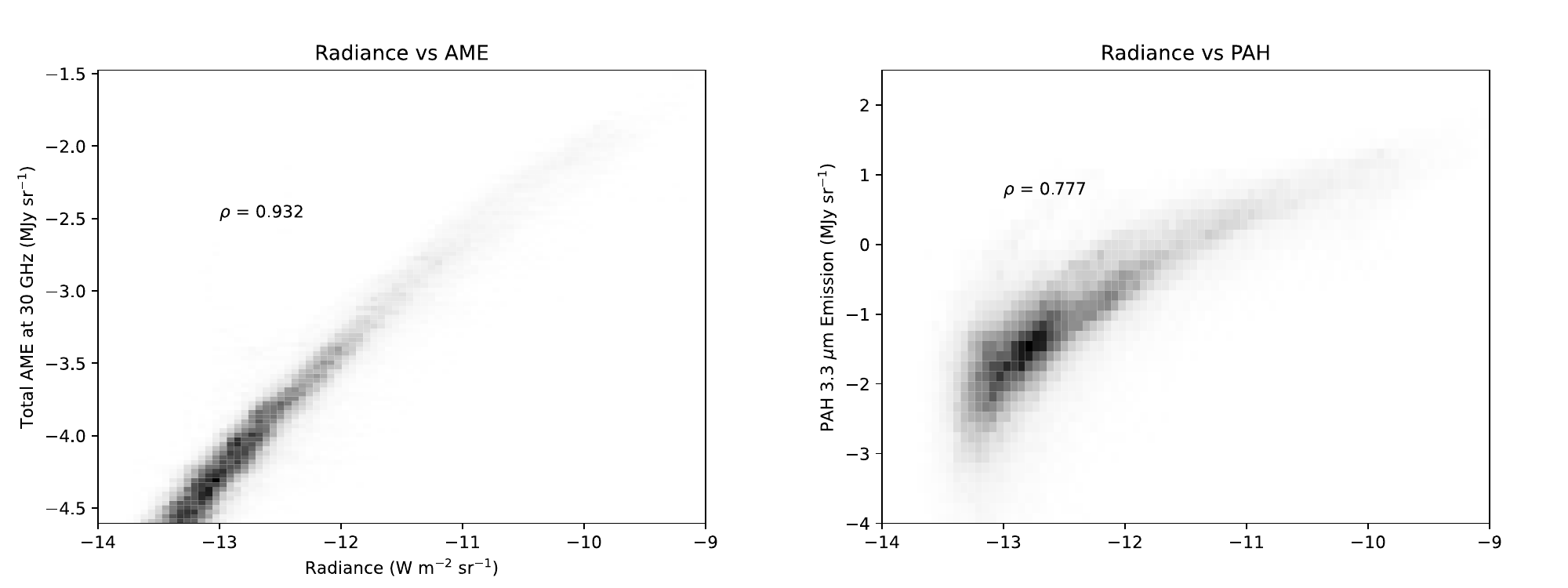}
    \caption{The correlations between the AME and PAH emission and the dust radiance from Planck. Left: AME vs radiance. Right: PAH vs radiance. Spearman-r correlation coefficients are shown in the upper-left of both panels.}
    \label{fig:rad_corr}
\end{figure*}
The AME is better correlated with the dust radiance than the PAH emission as can be seen in Figure \ref{fig:rad_corr}. We note, however, that the radiance is well-correlated with both the AME ($\rho$ = 0.93) and the PAH emission ($\rho$ = 0.78), indicating that the radiance is a good tracer of both the AME and PAHs. Our results also agree with previous studies indicating that the radiance is a better tracer of the AME than the PAH or far-infrared dust grain emission \citep{Hensley2016}.

\section{DISCUSSION} \label{sec:disc}
\begin{figure}
    \centering
    \includegraphics[width=0.45\textwidth]{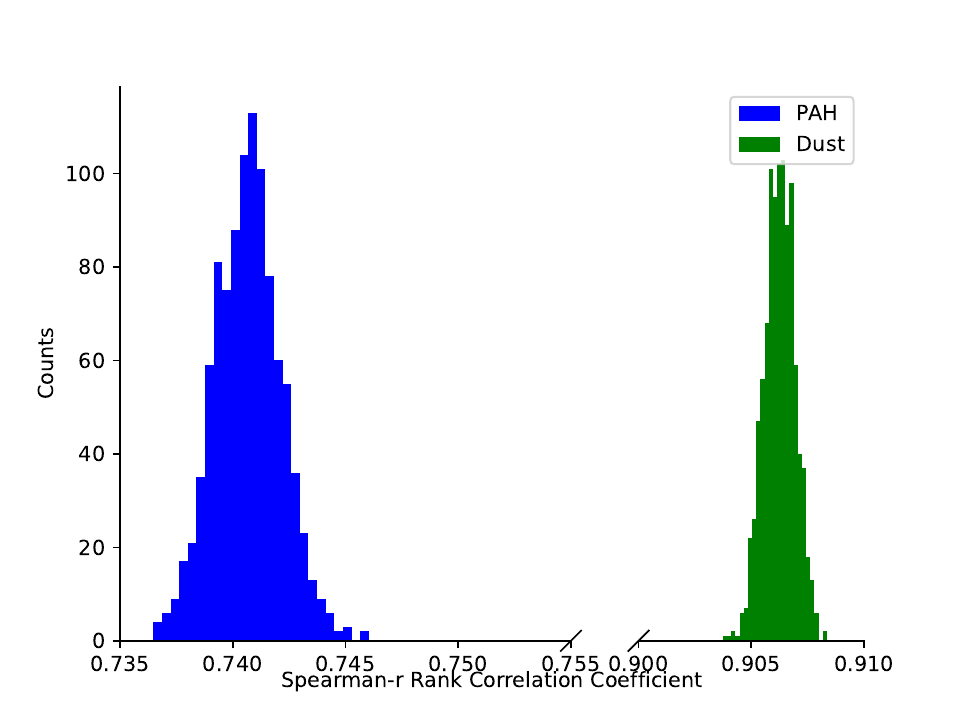}
    \caption{Bootstrap resampling results to estimate the uncertainty in the Spearman-r values obtained from the correlations shown in Figure \ref{fig:corr_full}. These distributions were obtained using 1000 samples of the larger distributions.}
    \label{fig:boot_r}
\end{figure}
\subsection{Milky Way AME Correlations with PAH and Dust Emission}
The AME is better correlated with the 857 GHz dust emission from Planck than it is for the PAH 3.3 $\mu$m emission feature derived from COSMOGLOBE DIRBE. This can be seen in Figure \ref{fig:corr_full} where the correlations and associated Spearman-r coefficients ($\rho=0.906$ and $\rho=0.740$) are displayed. Previous studies of targeted molecular structures show this same trend where the AME is better correlated with the dust emission than it is with the PAH emission \citep{Chuss2022, Sponseller2025}. The results presented in this work indicate that the AME throughout the Galactic Plane is generally better traced by the dust emission, agreeing with what is observed in the majority of compact AME sources studied previously.

We note, though, that 17 of the 98 prominent AME sources studied in \citet{Sponseller2025} show the opposite trend where they are better correlated with the PAH emission than the dust emission. In inspecting the distribution of these 98 sources we note that the majority of the preferentially PAH sources are at latitudes greater than 10\degree, and so therefore are not analyzed in this work. Only seven of the preferentially PAH sources are within the central 10\degree\ latitude region of the sky analyzed in this work, compared to $\sim$50 sources that are in this latitude range that are preferentially traced by dust. Our results therefore agree with those of \citet{Sponseller2025}, in that the bulk of the AME in the Galactic Plane is better traced by the dust emission.

To determine the significance of dust being the preferred AME tracer we calculate the uncertainties on the Spearman-r rank correlation coefficients for the dust and PAH correlations shown in Figure \ref{fig:corr_full}. We use a bootstrap method to do so where we sample with replacement the correlations and determine the Spearman-r coefficient for these samples. To perform this bootstrapping we used the pymccorrelation package\footnote{https://github.com/privong/pymccorrelation}, which allows for computation of correlation metrics like the Spearman-r and Pearson-r coefficients \citep{Curran2014,Privon2020}. We obtain distributions of Spearman-r coefficients from this method that are approximately Gaussian distributed. We are therefore able to derive the standard deviations of these Spearman-r distributions, $\sigma_{\rm PAH}$ and $\sigma_{dust}$. From this method we obtain uncertainties of $\sigma_{\rm PAH} = 1.4\times{}10^{-3}$ and $\sigma_{dust} = 7.4\times{}10^{-4}$.

We also display the Spearman-r distributions obtained from this bootstrap method to determine whether the distributions overlap in Spearman-r space. These distributions are shown as histograms in Figure \ref{fig:boot_r}. The bootstrap distribution for the AME vs dust correlation is at a significantly higher Spearman-r than the AME vs PAH distribution and these distributions do not overlap on the Spearman-r axis. The lack of any overlap between these distributions indicates that the preferential correlation for the AME and dust emission is highly significant.

To quantitatively determine whether the preferential correlation is significant we use the following equation:
\begin{equation}
    \eta_{pref} = \frac{\rho_{\rm PAH} - \rho_{dust}}{\sqrt{\sigma_{\rm PAH}^2 + \sigma_{dust}^2}}, \label{eq:pref}
\end{equation}
\\
where $\rho_{\rm PAH}$ and $\rho_{dust}$ are the correlation coefficients obtained for the PAH and dust correlations, respectively. The magnitude of $\eta_{pref}$ indicates the significance of the preference, with a negative $\eta_{pref}$ value indicating a that the dust is the preferred tracer of AME. We find $\eta_{pref} = -101$ for the correlations shown in Figure \ref{fig:corr_full}, corroborating the conclusion that there is a significant preference for the far-infrared dust emission as the primary tracer of the AME.

We also study how the correlations vary with Galactic latitude. Figures \ref{fig:PAH_corr} and \ref{fig:dust_corr} show the AME correlations with PAH and dust emission, respectively. Both panels of Figures \ref{fig:PAH_corr} and \ref{fig:dust_corr} indicate the Spearman-r coefficient for that correlation. The dust correlation remains significant ($\rho \geq 0.8$) for both latitude wedges, whereas the PAH correlation is only significant within the central 5.0\degree\ latitude wedge.

Though the PAH and dust emission correlations are both significant in the 5.0\degree\ regime, the dust correlation is more significant in this regime ($\rho=0.93$ for the dust correlation compared to $\rho=0.77$ for the PAH correlation). Furthermore, the Spearman-r coefficient is higher for the dust correlation than it is for the PAH correlation for both latitude regimes, as shown in Table \ref{tab:alignment}.

The rapid decrease in PAH correlation significance with latitude is perhaps explained by how the significance of the AME, PAH, and dust emission all decreases as $|b|$ increases. In particular, at high latitude regimes there is not much significant PAH emission as can be seen in Figure \ref{fig:PAH}, which is why we do not evaluate correlations for latitude magnitudes greater than 10\degree. This represents a current limitation of the DIRBE observations which future full-sky observations, such as those provided by SPHEREx, will improve upon \citep{Bock2025}.

We also study how the significance of the tracer (as quantified by $\eta_{pref}$) varies as a function of latitude. The change in $\eta_{pref}$ is shown in the bottom row of Table \ref{tab:alignment} for the two latitude wedges described above. We see that $\eta_{pref}$ becomes increasingly negative as latitude increases, indicating an increased preference of the far-infrared dust emission as the better tracer of the AME than the PAH emission. As mentioned previously, this increased preference is not dependent on the SNR of the PAH emission.

It is important to note that the 3.3 $\mu$m PAH emission feature studied here is only emitted by the smallest PAH grains. However, the smallest grains (PAH or non-PAH) are likely the dominant producer of the AME \citep[e.g.,]{Draine1998,Hoang2010}. Though the the larger PAH grains that are not traced by the 3.3 $\mu$m PAH emission feature, it is not likely to be a significant contributer to the observed AME.

\section{CONCLUSIONS} \label{sec:conc}
In this work, we have used Planck and COSMOGLOBE DIRBE full-sky maps to study whether the PAH or dust emission is a better tracer  of the AME in the Galactic Plane. We have found that, as in previous studies, the AME is better correlated with the dust emission than with the PAH emission. The analysis presented here is focused on the Galactic Plane ($|b| <$ 10\degree) since these lower latitudes are where Planck and DIRBE are not sensitivity limited.

The fact that the AME is better correlated with the dust emission than the PAH emission indicates that PAHs may not be the most likely explanation for the AME. Alternative mechanisms for AME have been proposed which may be more likely than PAHs. For example, spinning nano-silicate, non-PAH grains may have sufficient abundance to account for the AME \citep{Hoang2016a,Hensley2017}. Furthermore, spinning grains with magnetic dipoles could partially account for the observed AME \citep{Hoang2016b,Hensley2017}. Future work is needed to refine our understanding of which specific mechanism is primarily responsible for the AME. Studying the polarization of the AME, for example, can help determine the size distribution of the grains producing the AME. Smaller grains can be easily knocked out of alignment, meaning small grains are not expected to produce significant polarization. Previous studies of the AME polarization reveal low polarization fractions of $\leq$5\% \citep{Gonzalez-Gonzalez2025}, indicating that small grains like PAHs are the most likely dust population responsible for the production of AME.

An important caveat to this conclusion, however, is that PAH emission physics and AME have different dependencies on local interstellar conditions \citep{Hensley2022,Ysard2022}. This different dependency could erode the correlation between the two emission signatures, even if they originate from the same grains. We therefore cannot fully rule out the possibility that PAHs are the origin of the AME based on the results presented here.

The results presented here will be enhanced by the SPHEREx observatory, which will provide a direct observation of the the PAH emission thanks to its high spectral resolution \citep{Crill2020}. Furthermore, SPHEREx will provide an improved angular resolution (6'') and sensitivity ($>$19.4 AB mag for point sources) across the whole sky compared to Planck\footnote{https://spherex.caltech.edu/page/instrument}. The analysis presented here reveals that the properties of the diffuse emission at higher latitudes remains elusive. Follow-up work enhancing our understanding of the fainter, diffuse emission at these higher latitudes will be highly informative for determining which mechanisms are primarily responsible for the AME.

We summarize the key findings of this paper here:
\begin{itemize}
    \item We find that the AME is better correlated with the dust emission than the PAH emission in the Galactic Plane as seen in Figure \ref{fig:corr_full}. This result agrees with previous studies targeting specific molecular clouds that largely find that the AME in specific clouds is better correlated with the dust emission than the PAH emission \citep[e.g.,][]{Sponseller2025,Chuss2022}.
    \item The far-infrared dust emission is a better tracer of AME within a 10\degree\ latitude range of the Galactic Plane (as shown in Figures \ref{fig:dust_corr} and \ref{fig:PAH_corr}). The preference is significant as quantified by both the Spearman-r rank correlation coefficient and the $\eta_{pref}$ calculated for the correlations (both of which are presented in Table \ref{tab:alignment}).
    \item We find that at higher latitudes both Planck and DIRBE struggle to recover the fainter emission from this region of the sky. We are therefore not able to confidently determine the properties of the AME above a latitude of $|b|\sim10$\degree.
    \item Our results indicate that spinning PAHs may not be the most likely mechanism to explain the AME. Rather, mechanisms like spinning non-PAH nano-silicates or thermal vibrational emission might be more significant mechanisms producing the AME studied in this work \citep{Hoang2016a, Hensley2017}. Future full-sky work using instruments like SPHEREx will enhance our understanding of this point.
\end{itemize}

\begin{acknowledgements}
    We would like to thank the anonymous referee for their helpful comments on this work. This work was funded by NASA ADAP award number 80NSSC24K0623. This research was carried out in part at the Jet Propulsion Laboratory, California Institute of Technology, under a contract with the National Aeronautics and Space Administration. Some of the results in this paper have been derived using the healpy and HEALPix packages.
\end{acknowledgements}

\facility{
    DIRBE,
    Planck
    }

\software{
    Astropy \citep{Astropy2022,Astropy2018,Astropy2013},
    Matplotlib \citep{Hunter2007}, 
    Numpy \citep{Harris2020},
    PyMC,
    Healpy and HEALPix \citep{Zonca2019,Gorski2005}
    }

\appendix
\counterwithin{figure}{section}

\renewcommand{\thefigure}{A.\arabic{figure}}

\section{FSM Coefficient and Covariance Between PAH and FSM Coefficients} \label{sec:appendix}
\begin{figure*}
    \centering
    \includegraphics[width=1.0\textwidth]{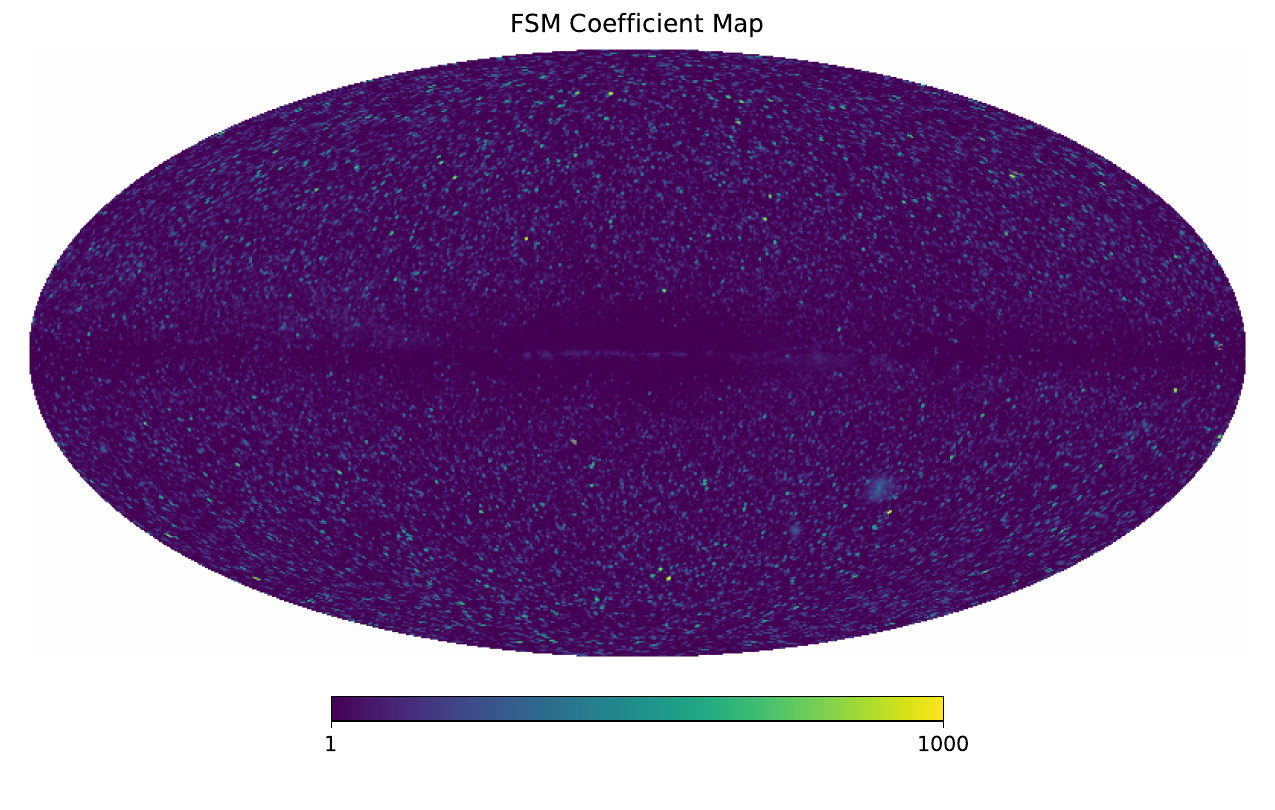}
    \caption{The FSM model coefficients obtained from the LLS model fitting of bands 1 -- 4 of the COSMOGLOBE re-reduction of the DIRBE observations. This coefficient corresponds to $a_2$ of Equation \ref{eq:Y} and is displayed in logarithmic scale.}
    \label{fig:FSM}
\end{figure*}
\begin{figure*}
    \centering
    \includegraphics[width=1.0\textwidth]{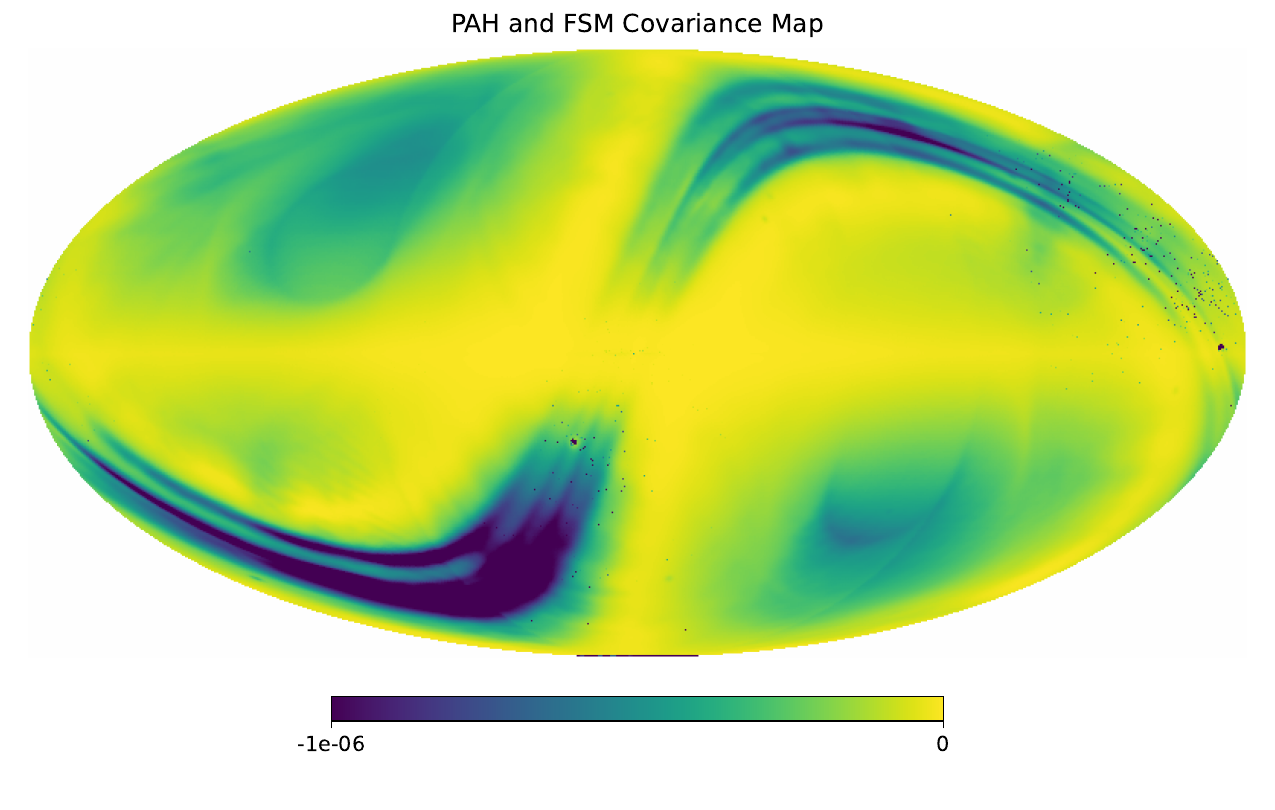}
    \caption{Covariance of the $a_1$ and $a_2$ coefficients from the LLS model fitting of the first four COSMOGLOBE DIRBE bands, where the $a_1$ coefficient is the PAH emission and the $a_2$ coefficient is the FSM emission. The covariance is displayed in a linear scale.}
    \label{fig:covar}
\end{figure*}
Figure \ref{fig:FSM} displays the model coefficient corresponding to the FSM emission of the LLS fitting method of the first four DIRBE bands as described in Section \ref{sec:meth}. Figure \ref{fig:covar} displays the covariance between the PAH ($a_1$) and FSM ($a_2$) coefficients obtained from the LLS model fitting. The covariance is predominantly close to zero indicating there is generally no relationship between the PAH and FSM model components.

\bibliography{PAH_III}{}
\bibliographystyle{aasjournal}

\end{document}